\documentclass[aps,prl,twocolumn,showpacs,notitlepage,floatfix,longbibliography,superscriptaddress,10pt]{revtex4-1}

\usepackage{amssymb}
\usepackage{amsmath}
\usepackage{graphicx}
\usepackage{braket}
\usepackage[colorlinks=true,linkcolor=blue,citecolor=blue,filecolor=blue,urlcolor=blue,pdfstartview=FitH,breaklinks=true]{hyperref}
\usepackage{xcolor}
\usepackage{ulem}
\usepackage{diagbox} 

\newcommand{\pksadd}{Max Planck Institute for the Physics of Complex Systems, Dresden D-01187, Germany}
\newcommand{\pcsadd}{Center for Theoretical Physics of Complex Systems, Institute for Basic Science(IBS), Daejeon 34126, Korea}
\newcommand{\ustadd}{Basic Science Program(IBS School), Korea University of Science and Technology(UST), Daejeon 34113, Korea}
\newcommand{\nziasadd}{New Zealand Institute for Advanced Study, Massey University, Auckland, Private Bag 102904, 0632 Auckland, New Zealand}

\newcommand{\sect}[1]{\textsl{#1} ---}

\newcommand{\mh}{\mathcal{H}}
\newcommand{\mhsp}{\ensuremath{\hat{\mh}_\text{sp}}}
\newcommand{\mhi}{\ensuremath{\hat{\mh}_\text{int}}}

\newcommand{\ha}{\hat{a}}
\newcommand{\hb}{\hat{b}}
\newcommand{\hc}{\hat{c}}
\newcommand{\hd}{\hat{d}}

\newcommand{\hf}{\hat{f}}
\newcommand{\hg}{\hat{g}}

\newcommand{\hn}{\hat{n}}

\newcommand{\ho}{\hat{I}}

\begin{document}

\title{Many-Body Flatband Localization}

\author{Carlo Danieli}
\affiliation{\pksadd}

\author{Alexei Andreanov}
\affiliation{\pcsadd}
\affiliation{\ustadd}

\author{Sergej Flach}
\affiliation{\pcsadd}
\affiliation{\ustadd}
\affiliation{\nziasadd}

\date{\today}

\begin{abstract}
    We generate translationally invariant systems exhibiting many-body localization from All-Bands-Flat single particle lattice Hamiltonians dressed with suitable short-range many-body interactions. This phenomenon -- dubbed Many-Body Flatband Localization (MBFBL) -- is based on symmetries of both single particle and interaction terms in the Hamiltonian, and it holds for any interaction strength. We propose a generator of MBFBL Hamiltonians which covers both interacting bosons and fermions for arbitrary lattice dimensions, and we provide explicit examples of MBFBL models in one and two lattice dimensions. We also explicitly construct an extensive set of local integrals of motion for MBFBL models. Our results can be further generalized to long-range interactions as well as to systems lacking translational invariance. 
\end{abstract}

\maketitle

\sect{Introduction}
Understanding the lack of thermalization in quantum interacting systems has been an active topic since Anderson predicted in 1958 the absence of transport in single particle lattices due to spatial  disorder~\cite{anderson1958absence}. This localization phenomenon has been extensively studied theoretically and experimentally~\cite{kramer1993localization}, with the impact of interaction between localized particles as one of the main open questions. Weak interactions were predicted to preserve the absence of transport of interacting particles~\cite{basko2006metal,aleiner2010finite} about fifty years after Anderson original work, leading to the phenomenon of \textit{Many-Body Localization} (MBL). The study of MBL systems and their properties is nowadays a very active topic of research with several open issues and active fronts - for a survey of the state of the art, see~\cite{abanin2017recent,abanin2019colloquium}.

MBL was initially predicted for interacting disordered systems emerging as an interplay of disorder and weak interactions. However it was later realized that the presence of disorder is not essential, launching the search for disorder-free MBL systems. Several possible scenarios emerged as a result: from non-ergodic behavior in networks of Josephson  junctions~\cite{pino2016nonergodic} to 1D fermionic lattices involving different species of particles~\cite{schiulaz2015dynamics} or the presence of d.c. field~\cite{schulz2019stark}, local constraints due to gauge invariance~\cite{karpov2020disorderfree}, presence of a large number of conserved quantities~\cite{smith2017disorder,smith2018dynamical}, quasi-periodic long-range interactions~\cite{mondaini2017many}, among others. Some proposals also explored the connection to glasses, predicting MBL in glassy systems~\cite{laumann2014many,baldwin2015the,baldwin2017clustering,mossi2017on}, {\it e.g.} kinetically constrained models~\cite{vanhorssen2015dynamics} and geometrically frustrated models~\cite{zhao2020glass}. However, the validity of some of the proposals were later doubted, as it was shown that several disorder-free MBL systems rely on vastly different energy scales and finite-size constraints~\cite{papic2015many}. In other cases instead ({\it e.g.}~\cite{pino2016nonergodic}), disorder-free MBL requires high temperatures or specific strong interaction regimes, likewise the original MBL requests weak interaction regimes.

In this letter, we propose a generator of disorder-free MBL systems which is free of the above-mentioned requirements (specific interaction or temperature regimes, finite-size constraints, type of many-body statistics, among others) and applies for arbitrary spatial dimensions. This generator relies on geometrical frustration of the translationally invariant single particle Hamiltonians which yields no single particle dispersion - i.e. all Bloch bands are dispersionless (or flat) - and suitably chosen many-body interactions. The resulting models exhibit non-ergodic behavior with lack of transport of particles for any interaction strength, and this phenomenon is dubbed {\it Many-Body Flatband Localization} (MBFBL). The study of networks with one or several flatbands (FB) is an active topic of research on its own. They were first discussed in the context of groundstate ferromagnetism~\cite{mielke1993ferromagnetism}, but were later identified in various other systems~\cite{leykam2018artificial,leykam2018perspective} and they have been experimentally realized in several setups, using e.g. ultra cold atoms~\cite{taie2015coherent} and photonic lattices~\cite{mukherjee2015observation,vicencio2015observation,weimann2016transport}. An important property of FB systems is the presence of compact localized states (CLS) - eigenstates with strictly finite support. These were used to systematically construct FB models~\cite{maimaiti2017compact,maimaiti2019universal,maimaiti2020flatband} along with other methods~\cite{flach2014detangling,dias2015origami,ramachandran2017chiral,roentgen2018compact,toikka2018necessary}. Their fine-tuned character makes FB systems an ideal platform to study diverse localization phenomena in the presence of onsite disorder~\cite{leykam2013flat,bodyfelt2014flatbands,danieli2015flatband,leykam2017localization}, DC fields~\cite{kolovsky2018topological}, and nonlinearities~\cite{danieli2018compact,ramachandran2018fano}, among many others. 

We introduce MBFBL networks formed by single particle All-Bands-Flat lattice Hamiltonians dressed with suitable short-range many-body interactions, and provide explicit examples in one and two spatial dimensions. We also discuss distinct interaction terms (including long-range interactions) in order to cover different types of particle statistics. We construct an extensive set of local integrals of motion present in MBFBL networks, and explicitly derive these integrals for some of the examples presented. We extend our generator scheme by removing the assumption of translation invariance of the lattice. 

\sect{Setup}
We consider a translationally invariant many-body Hamiltonian $\hat{\mh}$ on a lattice 
\begin{gather}
    \label{eq:Ham1}
    \hat{\mh} = \mhsp + \mhi\;,\qquad \mhsp = \sum_k \hf_k\;,\; \; \mhi = \sum_\kappa \hg_\kappa
\end{gather}
with both single particle part $\mhsp$ and interaction $\mhi$ written as sums of local operators $\hf_k$ and $\hg_\kappa$. The integers $k$ and $\kappa$ label unit cells of the lattice in a direct space for two different unit cell choices $A$ and $B$. We assume that the sites from one unit cell of e.g. choice $A$ belong to {\it different} unit cells of choice $B$. Regardless of the choice, each unit cell contains $\nu$ lattices sites or single particle levels. The operators are expressed through creation and annihilation operators $\hc_{k,a}^\dagger, \hc_{k,a}$ which create or annihilate a single particle on a given lattice site $k,a$ with $1\leq a \leq \nu$. Then the local operators read 
\begin{gather}
    \hf_k = \sum_{a,b=1}^{\nu} t_{ab} \hc_{k,a}^\dagger \hc_{k,b} + \text{h.c.}
    \label{eq:fk}
\end{gather}    
We assume the \textit{interaction} Hamiltonian $\mhi$ to be two-body, so that the local operators are
\begin{gather}
    \hg_\kappa = \sum_{\alpha,\beta,\gamma,\delta=1}^\nu J_{\alpha \beta \gamma \delta} \hc_{\kappa,\alpha}^\dagger \hc_{\kappa,\beta} ^\dagger \hc_{\kappa,\gamma} \hc_{\kappa,\delta} + \text{h.c.}
    \label{eq:gk}
\end{gather}

By the above definitions both single particle and interaction Hamiltonians are {\it semi-detangled} (SD) as $[ \hf_k , \hf_{k'} ] = [ \hg_{\kappa} , \hg_{\kappa'}] =0$ for any $k,k',\kappa ,\kappa '$. The spectrum of the single particle eigenvalue problem with $\mhsp$ yields $\nu$ flatbands with each being an eigenenergy of any of the local operators $f_k$. It follows that $\mhsp$ enforces full localization and absence of transport. The same is true for $\mhi$. However, because of the different unit cell choices $A,B$, in general it follows that $[\hf_k, \hg_\kappa]\neq 0$ for any given $k$ and at least a pair of different values of $\kappa$ (and vice versa). Consequently, the combination of both $\mhsp$ and $\mhi$ into $\mh$ in general yields transporting many-body eigenstates~\cite{vidal1998aharonov,vidal2000interaction,Tilleke2020nearest,danieli2020cagingprep,*danieli2020cagingprepII}.  

If $t_{ab}=t_{aa} \delta_{a,b}$ (with the Kronecker symbol $\delta_{a,b}$), the $\mhsp$ is coined {\it fully detangled} (FD)~\cite{flach2014detangling} since it depends on the particle number operators $\hn=\hc^\dagger \hc$ only, and does not move any particles from any lattice site to any other one. Together with $\mhi$ being SD, the full Hamiltonian $\mh$ preserves full localization of particles, which is an example of \textit{many-body flatband localization} (MBFBL). Likewise, if we assume that $J_{\alpha \beta \gamma \delta} = J_{\alpha \beta \alpha \beta} \delta_{\alpha,\gamma} \delta_{\beta,\delta}$ it follows that $\mhi$ is FD and does not move any particles from site to site. Together with $\mhsp$ being SD, we again arrive at the result that the full Hamiltonian $\mh$ lacks transporting eigenstates and is MBFBL. The relation between the FD/SD character of the Hamiltonians and the presence of MBFBL is summarized in Table~\ref{tab:H0H1}. We refer to all the other types of Hamiltonians as \textit{non-detangled} (ND).

\begin{table}[h]
    \centering
    \begin{tabular}{|c|c|c|}
        \hline
        \diagbox[width=3.75em]{$\mhsp$}{$\mhi$}
        & SD & FD\tabularnewline
        \hline
        SD  & --- &  MBFBL \tabularnewline
        \hline
        FD & MBFBL & MBFBL\tabularnewline
        \hline 
    \end{tabular}
    \caption{Existence of MBFBL for different types of single particle Hamiltonian $\mhsp$ and interaction Hamiltonian $\mhi$ as discussed in the main text.}
    \label{tab:H0H1}
\end{table}

We generate MBFBL Hamiltonians by choosing any of the FD/SD MBFBL models from Table~\eqref{tab:H0H1}. We then perform a unitary transformation (rotation) on each unit cell in either of the two unit cell choices $A, B$. This results in general in some complicated Hamiltonian with $\mhsp$ being ND and $\mhi$ being FD/SD, or vice versa - $\mhsp$ being FD/SD and $\mhi$ being ND - depending on which unit cell type the transformation was applied to. Furthermore these transformations can be chosen unit cell dependent resulting in non-translationally invariant Hamiltonians. 

Conventional disordered MBL systems are known to possess an extensive set of local integrals of motion~\cite{serbyn2013local,Ros2015integrals,abanin2017recent}, though explicit derivations are complicated. These integrals are used to explain relevant properties of these systems. Local integrals of motion can be explicitly derived for MBFBL networks.  With our proposed scheme and considering a SD single particle Hamiltonian $\mhsp$ in $\mh$~\eqref{eq:Ham1}, it follows that the expectation values of the operators $\ho_k = \sum_{a=1}^{\nu} \hn_{k,a}$  measure the number of particles in each local unit $\hf_k$ of $\mhsp$. These numbers are conserved in the presence of a FD interaction $\mhi$ (since $\mhi$ does not move particles from one to another site). It follows that each $\ho_k$ commutes with the full Hamiltonian.

The unitary transformations used to recast $\mhsp$ as ND yield $N$ local integrals of motion $\ho_k$ expressed in the new basis for the generated MBFBL lattice. The very same follows if a pair of FD single particle $\mhsp$ and a SD interaction $\mhi$ is picked from Table~\ref{tab:H0H1}. In this case, the operators $\ho_\kappa = \sum_{\alpha=1}^{\nu} \hn_{\kappa,\alpha} $ defined in each local unit $\hg_\kappa$ of $\mhi$ as well lead to $N$ local integrals of motion of the MBFBL lattice  after the unitary transformations have been applied. In the case of FD-FD Hamiltonians $\mhsp,\mhi$, the extensive set of local integrals of motion contains $\nu\times N$ elements, since each particle number operator $\hn_{k,a}$ commutes with the full Hamiltonian $\mh$.

Most of the generated MBFBL models, while being appealing from a mathematical point of view, could be hard to implement in experiments due to the complicated structure of the interaction $\mhi$ spanning several unit cells. Experimental feasibility instead favors fully detangled  $\mhi$, which result e.g. from Coulomb interactions between density operators in real space~\cite{ziman1972principles}. Therefore we refine our generator scheme by choosing  SD single particle $\mhsp$ and FD interaction $\mhi$, and recast $\mhsp$ to a ND Hamiltonian via unitary transformations that keep $\mhi$ fully detangled. This algorithm works for any number of bands $\nu$ of $\mhsp$, in any dimension, and any type of many-body statistics.

\sect{Results}
We will now discuss concrete examples in one and two spatial dimensions. We consider the SD Hamiltonian $\mhsp$ and conveniently restate it in the unit cell representation $B$ of $\mhi$. We then apply the subsequent unitary transformations. 
This change of unit cell introduces hopping terms between neighboring unit cells in each local Hamiltonian $\hf_\kappa$. 
Without loss of generality, we assume nonzero hoppings between nearest-neighboring unit cells only, and we adopt the conventions  used in Refs.~\cite{maimaiti2017compact,maimaiti2019universal} for flatband networks generators. Then a possible $D=1$ Hamiltonian $\mhsp$ reads
\begin{gather}
    \mhsp = \sum_\kappa \hf_\kappa = \sum_\kappa \left[\frac{1}{2}\hat{C}_\kappa^{\dagger T} H_0\hat{C}_\kappa + \hat{C}_\kappa^{\dagger T} H_1\hat{C}_{\kappa+1} + \text{h.c.}\right]
    \label{eq:H0}
\end{gather}
where we grouped the annihilation (creation) operators $\hc_{\kappa,a}$ ($\hc_{\kappa,a}^\dagger$) in $\nu$-dimensional vectors $\hat{C}_\kappa( \hat{C}_\kappa^\dagger)$. The matrices $H_0,H_1$ describe intra- and intercell hopping respectively,  and are chosen so as to enforce the SD condition $[\hf_\kappa,\hf_{\kappa'}] = 0$ for all $\kappa,\kappa'$. We remark that this $\mhsp$ is only one of the infinitely many realizations of a SD single particle Hamiltonian. 

The FD two-body interaction Hamiltonian $\mhi$ introduced above~(\ref{eq:Ham1},\ref{eq:gk}) is taken with the coefficients $J_{\alpha\beta\gamma\delta} = J_{\alpha\beta\alpha\beta} \delta_{\alpha,\gamma} \delta_{\beta,\delta}$ for each local component $\hg_\kappa$: $J_{\alpha\beta\alpha\beta}=1$ for $\alpha = \beta$ and $J_{\alpha\beta\alpha\beta}=2$ for $\alpha \neq \beta$. Then $\mhi$ is preserved as FD with the same coefficients $J_{\alpha\beta\gamma\delta}$ by any $2\times 2$ unitary transformation 
\begin{gather}
    U_{ab}:\
    \begin{cases}
        \hc_{\kappa,a} = z \hd_{\kappa,a} + w \hd_{\kappa,b}\\
        \hc_{\kappa,b} = -w^* \hd_{\kappa,a} + z^* \hd_{\kappa,b}\
    \end{cases}
    \label{eq:rot1}
\end{gather}
parameterized by two complex numbers $z,w$ such that $|z|^2 + |w|^2 = 1$ and any pair of sites $\hc_{\kappa,a}, \hc_{\kappa,b}$.

 The resulting Hamiltonian $\mhi$ for $\nu=2$ bands describes a two-body interaction among the sites $\ha_\kappa = \hc_{\kappa,a},\hb_\kappa = \hc_{\kappa,b}$ 
\begin{align}
    \mhi &= \sum_\kappa\bigl[\ha_\kappa^\dagger\ha_\kappa^\dagger\ha_\kappa\ha_\kappa + \hb_\kappa^\dagger\hb_\kappa^\dagger\hb_\kappa\hb_\kappa + 2\ha_\kappa^\dagger\ha_\kappa\hb_\kappa^\dagger\hb_\kappa\bigr] \notag\\
    &= \sum_\kappa \bigl[ \hn_{a,\kappa} +  \hn_{b,\kappa} - 1\bigr] \bigl[ \hn_{a,\kappa} +  \hn_{b,\kappa}\bigr] 
    \label{eq:H1_nu2}
\end{align}
with $\hn_{a,\kappa}=\ha_\kappa^\dagger\ha_\kappa$ and $\hn_{b,\kappa} = \hb_\kappa^\dagger\hb_\kappa$. We refer to such interaction as an \textit{extended Hubbard} interaction, which applies to both bosons and fermions with spin. 

\begin{figure}
    \includegraphics[width=0.925\columnwidth]{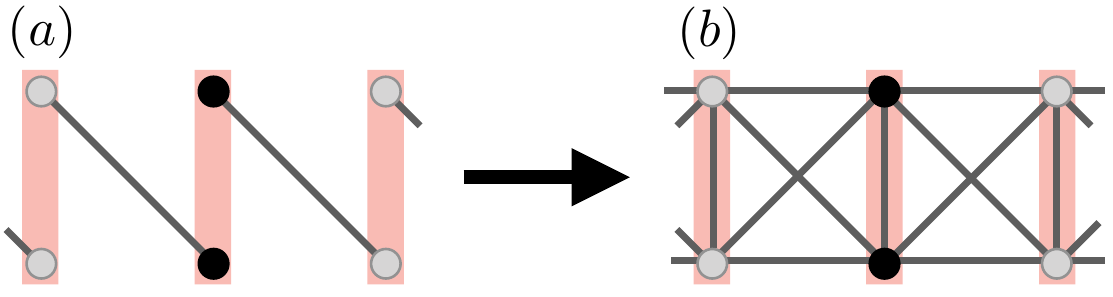}
    \caption{(Color online) One dimensional two band MBFBL network with $\mhsp$ SD (a) and with the cross-stitch lattice profile (b). The black circles indicate the unit cell choice, the solid lines correspond to sites connected by $\mhsp$ before (a) and after (b) the rotation, and the red shaded rods indicate the sites connected by the extended Hubbard terms~\eqref{eq:H1_nu2} of $\mhi$.}
    \label{fig:2bands}
\end{figure}

\sect{1D networks}
We now present two concrete examples of MBFBL networks. We first start with the simplest MBFBL network with $\nu=2$ bands. It is based on the Hamiltonian $\mhsp$ in Eq.~\eqref{eq:H0} with
\begin{gather}
    H_0 = \begin{pmatrix}
        1 & 0\\[0.3em]
        0 & 0
    \end{pmatrix},
    \qquad
    H_1 = \begin{pmatrix}
        0 & t\\[0.3em]
        0 &  0
    \end{pmatrix},
    \label{eq:H0H1_nu2}
\end{gather}
and a free complex parameter $t$. It is straightforward to check that this Hamiltonian is SD and has all bands flat. Next we pick the extended Hubbard interaction $\mhi$~\eqref{eq:H1_nu2}. The structure of $\mhsp$ and $\mhi$ is shown in Fig.~\ref{fig:2bands}(a) with solid lines and red shaded rods respectively. The rotation $U_{ab}$~\eqref{eq:rot1} recasts $H_0,H_1$~\eqref{eq:H0H1_nu2} as 
\begin{gather}
    H_0 = \begin{pmatrix}
        |z|^2 & -z w\\[0.3em]
        - z^* w^* & |w|^2
    \end{pmatrix},
    \quad
    H_1 = t 
    \begin{pmatrix}
        z w^* &  z^2\\[0.3em]
        - (w^*)^2 & -z w^*
    \end{pmatrix},
    \label{eq:H0H1_nu2_2}
\end{gather}
and makes Hamiltonian $\mhsp$ ND, while $\mhi$ remains FD. The resulting MBFBL network is shown in Fig.~\ref{fig:2bands}(b). The local integrals of motion read (after the rotation)
\begin{align}
   \ho_\kappa  &=
    |z|^2 ( \hat{n}_{a,\kappa-1} + \hat{n}_{b,\kappa}  ) 
+ |w|^2( \hat{n}_{a,\kappa} + \hat{n}_{b,\kappa-1}  ) \notag \\
&+ z^*w ( \hat{a}_{\kappa-1}^\dagger \hat{b}_{\kappa-1}  - \hat{a}_\kappa^\dagger\hat{b}_\kappa) + h.c. 
    \label{eq:d1nu2_liom}
\end{align}

\begin{figure}
    \includegraphics[width=0.975\columnwidth]{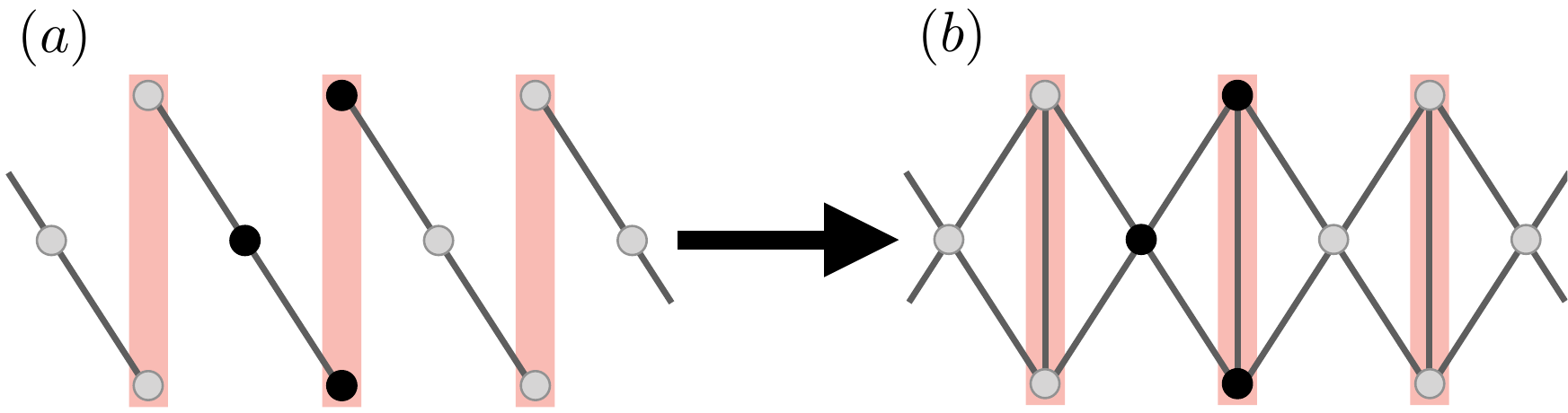}
    \caption{(Color online) One dimensional three band MBFBL network with $\mhsp$ SD (a) and with the diamond-shaped lattice profile (b). The black circles indicate the unit cell choice, the solid lines correspond to $\mhsp$ before (a) and after (b) the rotations, and the red shaded rods indicate the extended Hubbard terms~\eqref{eq:H1_nu2} of $\mhi$.}
    \label{fig:1D_3bands_DC}
\end{figure}

For three bands $\nu=3$ with operators $a_\kappa, b_\kappa, c_\kappa$ corresponding to the three sites of the unit cell, the SD Hamiltonian $\mhsp$~\eqref{eq:H1_nu2} has the following hopping matrices
\begin{gather}
    H_0 = 
    \begin{pmatrix}
        1 & 0 & 0 \\[0.3em]
        0 & 0 & t_1 \\[0.3em]
        0 & t_1^* & \mu
    \end{pmatrix},
    \qquad
    H_1 = 
    \begin{pmatrix}
        0 & 0 &t_2 \\[0.3em]
        0 & 0  & 0 \\[0.3em]
        0 & 0  & 0
    \end{pmatrix},
    \label{eq:H0H1_nu3}
\end{gather}
with two free complex ($t_1,t_2$) and one free real ($\mu$) parameters. This network is shown in Fig.~\ref{fig:1D_3bands_DC}(a) with gray solid lines. The interaction $\mhi$ consists of the extended Hubbard interaction~\eqref{eq:H1_nu2} between the top and the bottom sites $(a_\kappa,b_\kappa)$ of each plaquette (red shaded rods in Fig.~\ref{fig:1D_3bands_DC}(a)) and an additional optional onsite Hubbard interaction for the central site $c_\kappa$. Then the rotation $U_{ab}$~\eqref{eq:rot1} is applied to the pair $(a_\kappa,b_\kappa)$ only while leaving the sites $c_\kappa$ untouched. 
This recasts $H_0,H_1$~\eqref{eq:H0H1_nu3} into
%
\begin{gather}
    H_0 = 
    \begin{pmatrix}
       |z|^2 & -z w & t_1 w \\[0.3em]
        -z^* w^* & |w|^2 & t_1 z^* \\[0.3em]
        t_1^*w^* & t_1^* z & \mu
    \end{pmatrix}
    \quad
    H_1 = t_2
    \begin{pmatrix}
        0 & 0 & z \\[0.3em]
        0 & 0 & -w^* \\[0.3em]
        0 & 0 & 0
    \end{pmatrix}
    \label{eq:H0H1_nu3_2}
\end{gather}
defining a ND Hamiltonian $\mhsp$
while the interaction $\mhi$ remains FD. The resulting diamond-shaped MBFBL network is shown in Fig.~\ref{fig:1D_3bands_DC}(b). That diamond-shape profile has been realized in diverse experimental setups for flatband and compact localized state studies~\cite{mukherjee2015observation1,mukherjee2017observation,mukherjee2018experimental,xia2020observation,Kremer2020square}. Experimentally, the selective extended Hubbard interaction involving only the top and bottom sites  $\ha_\kappa,\hb_\kappa$ of the diamond plaquette might be achieved by reducing the distance between these sites as compared to the distance to the middle site $\hc_\kappa$. The parameter $t_1$ could be used to adjust the hoppings. 
The local integrals of motion $\ho_\kappa$ for this model are given by Eq.~\eqref{eq:d1nu2_liom} plus the additional particle number operator $\hn_{c,\kappa}$ for the central site $\hc_\kappa$ of the lattice, since it is unaffected by the rotation. 

\sect{2D networks} 
Construction of higher dimensional MBFBL networks follows a procedure similar to that of 1D systems.
In the simplest setting, the single particle Hamiltonian $\mhsp$ can be taken as a straightforward extension of Eq.~\eqref{eq:H0}, where matrices $H_1$ are replaced with matrices $H_0$ and $H_1^{(1)},\dots,H_1^{(D)}$ describing the intercell hopping along different spatial directions. The matrices are chosen to ensure that $\mhsp$ is SD. Now taking a suitable FD interaction $\mhi$, Eq.~\eqref{eq:H1_nu2} or its generalizations, and picking a unitary transformation that leaves this $\mhi$ FD, we obtain a ND Hamiltonian $\mhsp$. The full Hamiltonian $\mh$ exhibits MBFBL~\footnote{We point out that unlike 1D, the presence of several spatial directions may impose a constraint on the number of bands required to achieve MBFBL. For instance, the assumption of hopping between nearest neighbor unit cells in $\mhsp$ implies that 2D MBFBL networks have to have three of more bands.}.

\begin{figure}[!htbp]
    \centering
    \includegraphics[width=0.975\columnwidth]{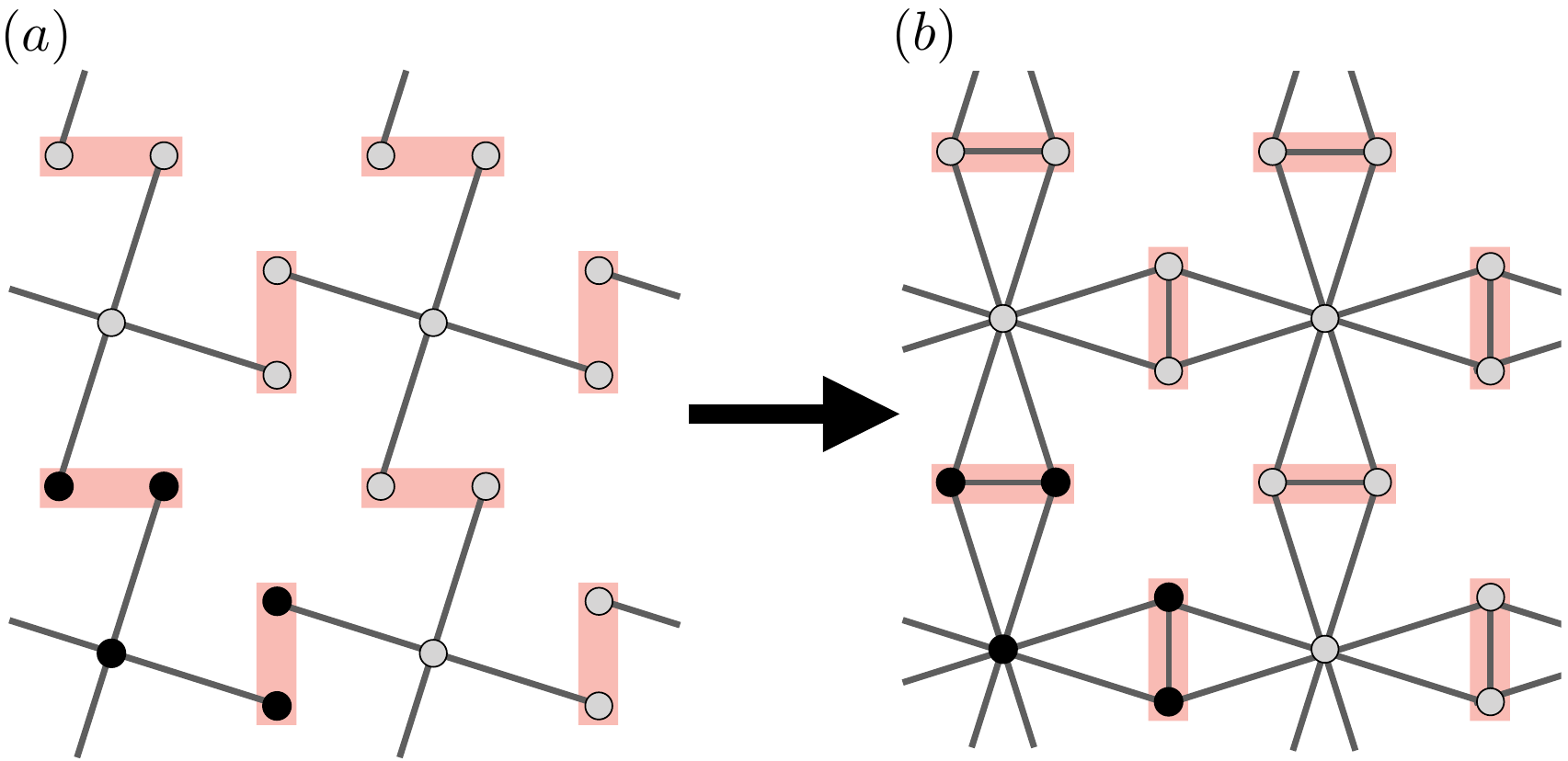}
    \caption{(Color online) Two dimensional five band MBFBL network with $\mhsp$ SD (a) and with the decorated Lieb lattice profile (b). The black circles indicate the unit cell choice, the solid lines correspond to $\mhsp$ before (a) and after (b) the rotations, and the red shaded rods indicate the extended Hubbard terms~\eqref{eq:H1_nu2} of $\mhi$.}
    \label{fig:2D_5bands}
\end{figure}

A notable two dimensional lattice exhibiting MBFBL obtained by applying these rules is the \textit{decorated Lieb} lattice~\cite{rontgen2019quantum}. This is a five-band $\nu=5$ network, whose SD Hamiltonian $\mhsp$ is shown in Fig.~\ref{fig:2D_5bands}(a), with matrices $H_0, H_1^{(1)}, H_1^{(2)}$. In each unit cell, we use the extended Hubbard Hamiltonians $\mhi$~\eqref{eq:H1_nu2} for the two site pairs indicated by red shaded rods in Fig.~\ref{fig:2D_5bands}(a), and an onsite Hubbard interaction for the central site. The two rotations $U_{ab}$~\eqref{eq:rot1} applied to the highlighted pairs (leaving the central site untouched) yield a ND $\mhsp$ shown in Fig.~\ref{fig:2D_5bands}(b), and the resulting full Hamiltonian $\mh$ is MBFBL. The local integrals of motion for the decorated Lieb lattice can be easily derived and have similar but more involved expressions to those of the previous models~\eqref{eq:d1nu2_liom}.

\sect{Perspectives}
The proposed scheme relies on the two-body Hamiltonian $\mhi$ with onsite terms in the interaction, restricting the interacting particles to bosons or spinful fermions. However, the same construction can be implemented for spinless fermions by \textit{e.g.} choosing local operators $\hg_{\kappa}^{\sigma} = \sum_{\alpha,\beta,\gamma,\delta=1}^{\nu} J_{\alpha \beta \gamma \delta}^{\sigma} \hc_{\kappa,\alpha}^\dagger \hc_{\kappa+\sigma,\beta} ^\dagger \hc_{\kappa,\gamma} \hc_{\kappa+\sigma,\delta} + h.c.$ with exclusively inter-site interaction terms between unit cell $\kappa$ and unit cell $\kappa+\sigma$. In particular, $\mhi$ is FD for $J_{\alpha\beta\gamma\delta}^\sigma=J \delta_{\alpha,\gamma} \delta_{\beta,\delta}$ and it is preserved as FD by the same transformation~\eqref{eq:rot1}. This yields a generator of $D$-dimensional $\nu$-band MFBFL lattices for spinless fermions, with the recent work of Kuno \textit{et al.}~\cite{kuno2020flat} being a particular $D=1$ $\nu=2$ band example. The construction can be further extended to long-range all-to-all interaction Hamiltonians $\mhi$ by setting $\hg_\kappa = \sum_\sigma v_\sigma \hg_\kappa^\sigma$ and even infinite-range interactions $\mhi=J/N\sum_{\kappa\neq\kappa', a} \hn_{\kappa, a} \hn_{\kappa',a}$. The latter example is valid because the interaction is a function of the total density $\hat\rho = \sum_{\kappa,a} \hn_{\kappa,a}$ only and is therefore invariant under the transformation~\eqref{eq:rot1}.

We note that it is possible to extend the generator by abandoning the translational invariance of the Hamiltonian $\mh$.
We can choose the hopping parameters $t_{ab}$ and the interaction matrix elements $J_{\alpha \beta \gamma \delta}$  in the starting Hamiltonians $\mhsp$ and $\mhi$ respectively to be unit cell dependent. To stick with the proposed scheme where $\mhi$ is FD and is preserved by unitary transformations~\eqref{eq:rot1}, the unit cell dependent terms are restrained to the SD $\mhsp$ only ({\it e.g.} onsite or hopping disorder). The unitary transformations~\eqref{eq:rot1} used to recast $\mhsp$ as ND induce correlations between the onsite energies of the pairs of sites involved. In the models presented - Figs.\ref{fig:2bands}(b), \ref{fig:1D_3bands_DC}(b), \ref{fig:2D_5bands}(b) - these correlations are between the sites within the same red-shaded area. These correlations depend of the parameters $z,w$ defining $U_{a,b}$ in Eq.~\eqref{eq:rot1}. These parameters may also be chosen to vary upon changing $\kappa$ if the unitary transformations considered differ from unit cell to unit cell. Let us additionally observe that the breaking of translation invariance does not destroy the existence of the extensive set of local integrals of motion - they are given by the same operators as in the translationally invariant case.

\sect{Conclusions} 
We have introduced a generator of Many-Body Localized disorder-free Hamiltonians by applying unitary transformations to suitably detangled Hamiltonians -- a feature that assumes all-band-flat single particle Hamiltonians. This new phenomenon -- coined Many-Body Flatband Localization -- implies strict localization of any number of particles irrespective of dimensionality or interaction strength, and it does not require vastly different energy scales similar some models supposed to exhibit disorder-free MBL. Our work substantially extends previous studies of localization phenomena of interacting quantum many-body platforms with All-Band-Flat lattice single particle Hamiltonians~\cite{vidal2000interaction,junemann2017exploring,mondaini2018pairing,murad2018performed,Barbarino2019topological,roy2019compact,kuno2020flat,orito2020exact,danieli2020cagingprep}. In particular, we propose a flexible and general set of many-body localized systems which may be experimentally feasible. A novel and unique feature of these systems is the existence of unitary mappings that recast them into a detangled form. This very property can be employed to study the impact of additional perturbations of the proposed networks which lift MBFBL and modify the proposed local integrals of motion in a systematic and analytical form. Hence, these systems offer innovative and powerful tools to potentially  perform systematic analytical studies of conventional properties of MBL networks which typically  relay on heavy numerical studies. 

\sect{Acknowledgments}
This work was supported by the Institute for Basic Science (Project number IBS-R024-D1).
SF acknowledges support by the New Zealand Institute for Advanced Study where part of this work was completed.

\normalem
\bibliography{general,flatband,mbl}

\end{document}